\documentclass[doublecol]{epl2}

\newcommand{\be}[1]{\begin{equation}\label{eq:#1}}
\newcommand{\ee}{\end{equation}}
\newcommand{\bea}{\begin{eqnarray}}
\newcommand{\eea}{\end{eqnarray}}
\newcommand{\rmi}{{\rm i}}
\newcommand{\fl}{}
\newcommand{\Tr}{Tr}
\newcommand{\tr}{tr}
\newcommand{\ack}{\acknowledgements}

\newcommand{\bi}{\mathbf}
\newcommand{\phd}{\phantom{\dag}}
\newcommand{\up}{^{\phd}}
\newcommand{\ph}{\phantom{.}}
\newcommand{\noi}{\noindent}
\newcommand{\no}{\nonumber}

\title{Spontaneous Quantum Hall Effect in chiral d-density waves}
\shorttitle{Spontaneous Quantum Hall Effect in chiral d-density
waves}

\author{Kotetes P.\thanks{E-mail: \email{pkotetes@central.ntua.gr}}
\and Varelogiannis G.\thanks{E-mail:
\email{varelogi@central.ntua.gr}}} \shortauthor{Kotetes P. \etal}

\institute{
  \inst{} Department of Physics, National Technical University
of Athens-GR-15780 Athens, Greece}

\pacs{73.43.-f}{Quantum Hall effects} \pacs{71.27.+a}{Strongly
correlated electron systems; heavy fermions}
\pacs{71.45.Lr}{Charge-density-wave systems }

\abstract{We study the electromagnetic response of a chiral ${\rm
d_{ xy}+id_{x^2-y^2}}$ density wave state. Due to parity (${\cal
P}$) and time reversal (${\cal T}$) violation, Chern-Simons terms
emerge in the effective action of the U(1) gauge field. As a
consequence electric and magnetic fields are coupled providing the
possibility of observing the Spontaneous Quantum Hall Effect i.e.
generation of Hall voltage via the sole application of an electric
field. We demonstrate how the Chern-Simons terms are induced and
discuss the topological origin of the quantization of the Hall
conductance.}

\begin{document}

\maketitle

Among the most fascinating examples of exotic ordered states that
one may observe in strongly correlated electronic systems, are
those that violate simultaneously parity $({\cal P})$ and time
reversal ${(\cal T)}$ i.e. possess {\it chirality}. So far, this
type of states has been extensively considered in the context of
unconventional superconductivity \cite{Sigrist1}. The first
example of a superfluid chiral ground state was the
$\rm{p_x+ip_y}$ state invoked first for $^3He$
\cite{LeggettReview,WolfleBook,VolovikHe,VolovikYakovenkoHe,GoryoHe,VolovikBook}
and later in the context of superconductivity in crystalline
materials like Ruthenates
\cite{Rice1,Goryo,GoryoVortex,{Furusaki},Read,{YakovenkoHall}}.
Another chiral superconducting state is the
$\rm{d_{x^2-y^2}+id_{xy}}$ state that was proposed especially for
high-$T_c$ superconducting cuprates
\cite{{Goryo},{Read},Sigrist,{Matsumoto},{Horovitz}}. The
coexistence of the two different order parameter components in the
above cases, is not only affecting the nodal structure of the
ground state, but has also extraordinary implications on their
electromagnetic behaviour.

The simultaneous ${\cal P-T}$ violation induces the so called
Chern-Simons terms
\cite{VolovikBook,Deser,Wilzec,Semenoff,Haldane,Fradkin,KerlerHall}
in the effective action of the electromagnetic field. The
connection of these terms to the Quantum Hall Effect (QHE) and
fractional statistics is well known
\cite{Fradkin,KerlerHall,KerlerC,Khare}. In the ordinary QHE,
these terms emerge because the magnetic field that we apply
perpendicular to the two dimensional electronic system, in the
presence of an in plane electric field, breaks chirality. In
addition, the topological structure of the emerging Chern-Simons
terms, causes the quantization of the Hall conductance
\cite{KerlerHall,TKNN,{Niu}}. In the case of chiral
superconductors, chirality is broken spontaneously giving rise
 to Chern-Simons terms leading to the so
called Spontaneous Quantum Hall Effect (SQHE)
\cite{VolovikBook,Goryo,Read,Horovitz}. This unusual
magneto-electric effect is generated because the electrons already
feel an effective magnetic field in the ground state, coming from
the angular momentum of the cooper pairs. Detailed studies of the
SQHE for both $\rm{p_x+ip_y}$ and $\rm{d_{x^2-y^2}+id_{xy}}$
chiral superconducting states are already available
\cite{VolovikBook,Goryo,GoryoVortex,Furusaki,Read,Horovitz}.

In this paper we consider instead a chiral ground state which is
{\it not} superconducting. Particularly, we study the
electromagnetic properties of the commensurate chiral
$\rm{d_{xy}+id_{x^2-y^2}}$ density wave (CDDW)
\cite{Wen,Yakovenko}. Unconventional spin singlet density wave
states \cite{Gruner,Berlinsky,Schulz,Thalmeier,Nayak1} have been
considered as potentially relevant situations in virtually all
strongly correlated systems of interest. In particular it has been
suggested that the $\rm{id_{x^2-y^2}}$ density wave state (DDW),
also called orbital antiferromagnetic state, may be relevant for
high-$T_c$ cuprates \cite{Nayak2} as well as in numerous heavy
fermion systems like URu$_2$Si$_2$ and CeCoIn$_5$ \cite{Dora} and
in organic metals like e.g. $\alpha$-BEDT salts. An
$\rm{id_{x^2-y^2}}$ density wave does not lead to a fully gapped
ground state as it leaves four nodes on the diagonals of the
Brillouin zone at the points
$\left(\pm\frac{\pi}{2},\pm\frac{\pi}{2}\right)$. Of course, the
system would prefer to gap these nodal points so to obtain a fully
gapped spectrum possibly by developing another density wave
component which will gap the nodal points. A natural candidate
order parameter for this situation is a $\rm{d_{xy}}$ density
wave. However, a $\rm{d_{xy}}$ component cannot be realized in the
half-filled case. The reason is that a $\rm{d_{xy}}$ order
parameter violates parity in 2-dimensions while the underlying
system preserves it, as the points $\left(\pm\pi,0\right)$ and
$\left(0,\pm\pi\right)$, related by reflections, are equivalent
since their difference is a reciprocal lattice vector. Still, this
symmetry argument can be surpassed as it does not stand if we
depart even slightly from half-filling. This is exactly the case
we consider. Specifically, we consider that our system is very
close to half-filling so as to be favourable and permissible to
develop a composite density wave state of the form
$\rm{d_{xy}+id_{x^2-y^2}}$ of wave-vector
$\bi{Q}=\left(\pi,\pi\right)$.

Based on a microscopic model under the above assumptions, we
demonstrate how the ${\cal P-T}$ violation induces the
Chern-Simons terms in the effective action of the electromagnetic
field at one-loop level. We discuss the characteristics of the
SQHE that emerges and the concomitant topological quantization of
the Hall conductance. Although our study may be viewed as a
generalization to a chiral particle-hole condensate of previous
studies on chiral superconductors and indeed bears profound
similarities to the superconducting case, we should remark that it
is certainly not just a trivial extension of the former. In fact,
contrary to the chiral superconductors case, a commensurate chiral
d-density wave does not violate gauge invariance nor is
characterized by any Goldstone modes as translational symmetry is
not a continuous symmetry in this case \cite{Gruner,Berlinsky}.
Consequently, we expect no infrared dependence of the Hall
conductance \cite{GoryoHe}, while the arising spontaneous Hall
currents are locally conserved.



We consider a 2-dimensional tight binding model of interacting
one-band electrons on a square lattice close to half-filling. In
the following analysis, since we are interested only in spin
singlet density wave instabilities, we do not need to take into
consideration the electron spin, so we shall omit the spin
indices. Moreover, we consider the zero-temperature case and adopt
the following conventions: $k^i=\bi{k}=(k_{\rm x},k_{\rm y});\ph
k^{\mu}=k=(\omega,\bi{k});\ph k'=(\omega,\bi{k}');x^i=\bi{x}=({\rm
x,y});\ph x^{\mu}=x=(t,\bi{x});\ph i=1,2;\ph \mu=0,1,2;\ph
\hbar=1;\ph c=1;\int_{x}=\int{\rm d}{\rm t}{\rm d}^2{\rm
x};\ph\int_{\omega}=\int{\rm d}\omega/(2\pi);\int_q=\int{\rm
d}^3{\rm q}/(2\pi)^3$. We also consider that repeated indices are
summed.
\par Our starting point is the following action

\bea S&=&\int_{\omega}\sum_{\bi{k}}\ph
\bar{c}_{k}^{\phd}\left[\omega-\xi(\bi{k})\right]c_{k}^{\phd}\no\\&-&\frac{1}{2}\int_{\omega}\sum_{\bi{\ph
k,k'}} \ph \bar{c}_{k}^{\phd}c_{k+\bi{Q}}\up{\cal
V}(\bi{k,k'})\bar{c}_{k'+\bi{Q}}^{\phd}c_{k'}\up\,,\eea

\noi which contains an effective 2-body interaction ${\cal
V}(\bi{k,k'})\equiv {\cal V}(\bi{k,k'+Q,k+Q,k'})$. The Grassmann
variables $\bar{c}_{k}^{\phd},c_{k}^{\phd}$ correspond to the
electron creation and annihilation operators of frequency $\omega$
and wave-vector $\bi{k}\in 1^{st}$ Brillouin zone. The wave-vector
$\bi{Q}=(\pi,\pi)$ is the nesting wave-vector. The tight binding
energy dispersion $\xi(\bi{k})$ can be decomposed into periodic
and antiperiodic $\delta(\bi{k})$ and $\varepsilon(\bi{k})$ parts,
satisfying the relations

\bea \xi({\bi{k}})&=&-t(\cos k_{\rm{x}}+\cos
k_{\rm{y}})+t'\cos k_{\rm{x}}\cos k_{\rm{y}}\,,\\
\xi(\bi{k})&=&\varepsilon(\bi{k})+\delta(\bi{k})\,,\\
\varepsilon(\bi{k})&=&-\varepsilon(\bi{k+Q})\,,\label{eq:nesting}\\
\delta(\bi{k})&=&+\delta(\bi{k+Q})\,.\eea

\noi The antiperiodic part $\varepsilon(\bi{k})$ satisfies the
nesting condition expressed by eq.~(\ref{eq:nesting}), providing
tendency towards the formation of the density wave especially at
half-filling where this term dominates. Since in the situation
under consideration the system is very close to half-filling we
consider $\xi(\bi{k})\simeq\varepsilon(\bi{k})$ and neglect the
periodic part $\delta(\bi{k})$.

\par The interaction potential ${\cal V}$, is a
non retarded effective 2-body potential which drives the system
towards the formation of a density wave of the commensurate
($\bi{k}+2\bi{Q}=\bi{k}$) wave-vector $\bi{Q}=(\pi,\pi)$. In order
to decouple this action in the particle-hole channel, we execute a
Hubbard-Stratonovich transformation by inserting the auxiliary
fields $\Phi(k,k+\bi{Q}),\Phi(k+\bi{Q},k)$. This way we obtain the
following decoupled interaction part $\tilde{S}_{{\rm
int}}^{\phd}$

\bea \fl\tilde{S}_{{\rm
int}}&=&\frac{1}{2}\int_{\omega}\sum_{\phd\bi{k,k'}}\ph\Phi(k,k+\bi{Q}){\cal
V}(\bi{k,k'})\Phi(k'+\bi{Q},k')\no\\
&-&\frac{1}{2}\int_{\omega}\sum_{\phd\bi{k,k'}}\ph {\cal
V}(\bi{k,k'})\Phi(k'+\bi{Q},k')\bar{c}_{k}^{\phd}c_{k+\bi{Q}}^{\phd}\no\\
&-&\frac{1}{2}\int_{\omega}\sum_{\phd\bi{k,k'}}\ph {\cal
V}(\bi{k,k'})\Phi(k,k+\bi{Q})\bar{c}_{k'+\bi{Q}}^{\phd}c_{k'}^{\phd}\,.\eea

\noi At this point we define the density wave order parameters

\bea \Delta(k,k+\bi{Q})&=&\sum_{\bi{k}'}\ph{\cal
V}(\bi{k,k'})\Phi(k'+\bi{Q},k')\,,\\
\Delta(k+\bi{Q},k)&=&\sum_{\bi{k}'}\ph{\cal
V}(\bi{k',k})\Phi(k',k'+\bi{Q})\,.\eea

\noi The introduction of the order parameters simplifies further
the expression of $\tilde{S}_{{\rm int}}^{\phd}$

\bea\fl \tilde{S}_{{\rm
int}}^{\phd}&=&\frac{1}{2}\int_{\omega}\sum_{\bi{k}}\left.
\Phi(k,k+\bi{Q})\Delta(k,k+\bi{Q})\right.\no\\
&-&\frac{1}{2}\int_{\omega}\sum_{\bi{k}}\Delta(k,k+\bi{Q})\bar{c}_{k}^{\phd}c_{k+\bi{Q}}^{\phd}\no\\
&-&\frac{1}{2}\int_{\omega}\sum_{\bi{k}}
\Delta(k+\bi{Q},k)\bar{c}_{k+\bi{Q}}^{\phd}c_{k}^{\phd}\,.\eea

\noi The total action, $\tilde{S}$, describing the decoupled
system obtains the following matrix form

\bea \fl\tilde{S}&=&\frac{1}{2}\int_{\omega}\sum_{\bi{k}}
\left(\begin{array}{cc}\bar{c}_{k}^{\phd}&\bar{c}_{k+\bi{Q}}^{\phd}\end{array}\right)\no\\
&\times&\left(\begin{array}{cc}\omega-\varepsilon({\bi{k}})&-\Delta(k,k+\bi{Q})\\
-\Delta(k+\bi{Q},k)&\omega+\varepsilon({\bi{k}})\end{array}\right)
\left(\begin{array}{c}c_{k}^{\phd\ph}\phantom{\bi{Q}}\\
c_{k+\bi{Q}}^{\phd}\end{array}\right)\no\\
&+&\frac{1}{2}\int_{\omega}\sum_{\bi{k}}\Phi(k,k+\bi{Q})\Delta(k,k+\bi{Q})\,.\eea

\noi With the introduction of the isospin Pauli matrices
$\tau_1^{\phd},\tau_2^{\phd},\tau_3^{\phd}$

\bea \tau_1^{\phd}=\left(\begin{array}{cc}0&\ph 1\\1\ph&\ph
0\end{array}\right)\ph
\tau_2^{\phd}=\left(\begin{array}{cc}0&-\rmi\\\rmi&\ph
0\end{array}\right)\ph \tau_3^{\phd}=\left(\begin{array}{cc}1&\ph
0\\0&-1\end{array}\right)\eea

\noi and the following isospinor
$\bar{\Psi}_{k}^{\phd}=\frac{1}{\sqrt{2}}(\ph
\bar{c}_{k}^{\phd}\ph \bar{c}_{k+\bi{Q}}^{\phd})$, the total
action becomes

\bea\fl\tilde{S}&=&\int_{\omega}\sum_{\bi{k}}
\frac{1}{2}\Phi(k,k+\bi{Q})\Delta(k,k+\bi{Q})\no\\
&+&\int_{\omega}\sum_{\bi{k}}\bar{\Psi}_{k}^{\phd}\left[\omega-\bi{g}(k)\cdot
\bi{\tau}\right]\Psi_{k}^{\phd}\no\\&-&\int_{\omega}\sum_{\bi{k}}
\int_q\bar{\Psi}_{k+q}^{\phd}\ph\Gamma_{\mu}(k)A^{\mu}(q)\Psi_{k}^{\phd}\,,\eea

\noi where we have introduced the isospin vector $\bi{g}(\bm{k})$,
with components

\bea
{\rm g}_1\up(k)&=&+\left[\Delta(k,k+\bi{Q})+\Delta(k+\bi{Q},k)\right]/2\,,\\
{\rm
g}_2\up(k)&=&-\left[\Delta(k,k+\bi{Q})-\Delta(k+\bi{Q},k)\right]/2{\rm
i}\,,\\{\rm g}_3\up(k)&=&+\varepsilon(\bi{k})\,.\eea

\noi and the electron-photon interaction term. The interaction
vertices are defined as
\bea \Gamma_{0}^{\phd}(k)&=&-e\,,\\
     \Gamma_{i}^{(p)}(k)&=&+e\frac{\partial\bi{g}({k})}{\partial k^i}\ph \cdot\bi{\tau}\,,\\
     \Gamma_{i}^{(d)}(k)&=&-\frac{e^2}{2m}\ph A_{i}^{\phd}(-q)\,,\eea

\noi where the paramagnetic vertices $\Gamma_{i}^{(p)}$ are
obtained by the substitution $\bi{k}\rightarrow\bi{k}+e\bi{A}$,
justified on grounds of gauge invariance.
\par In order to proceed, first we
have to integrate the fermionic degrees of freedom at a stationary
solution of the fields $\Phi(\bi{k,k+Q})$, $\Phi(\bi{k+Q,k})$ and
$A_{\mu}(\bi{q})$. We consider that the U(1) currents are zero in
the ground state, implying the same for the gauge fields
$A_{\mu}$. Under these conditions, we obtain the ground state
action $\tilde{S}_{\rm{gs}}$ defined by the relation \bea
\tilde{S}_{\rm{gs}}=\frac{1}{2}\sum_{\bi{k}}
\Phi(\bi{k,k}+\bi{Q})\Delta(\bi{k,k}+\bi{Q})-\frac{\rmi}{2}\Tr
\ln\left[\hat{{\cal G}}^{-1}\right]\eea

\noi where we have introduced the operator version $\hat{{\cal
G}}$ of the one particle Green's function ${\cal G}$, satisfying
\bea {\cal G}^{-1}(k)=\omega-\bi{g}(\bi{k})\cdot
\bm{\tau}\label{eq:s}\,.\eea

\noi $\Tr$ denotes trace over wave-vector, frequency and isospin
variables. The factor $1/2$ is included in order to avoid double
counting because the variables $c_{k}^{\phd}$ and
$c_{k+\bi{Q}}^{\phd}$ are not independent. Minimization of the
ground state action with respect to the auxiliary fields $\Phi$,
yields the following self-consistence equation determining the
stationary value $\Delta(\bi{k})\equiv\Delta(\bi{k,k+Q})$

\bea\Delta(\bi{k})=-\rmi\int_{\omega}\sum_{\bi{k}'} {\cal
V}(\bi{k},\bi{k'})\tr\left[{\cal
G}(k')\frac{\tau_{1}^{\phd}-\rmi\tau_{2}^{\phd}}{2}\right]\,,\eea

\noi where $\tr$ denotes trace over isospin indices.


In order to obtain a deeper insight of the chiral d-density wave
characteristics, we examine the symmetry properties of the density
wave order parameters. First we define the order parameter matrix
in $\{k,k+\bi{Q}\}$ subspace

\bea {\cal
D}(k)=\left(\begin{array}{cc}0&\Delta(k,k+\bi{Q})\\
\Delta(k+\bi{Q},k)&0\end{array}\right)\,,\eea

\noi and introduce the translation operator $t_{\bi{Q}}^{\phd}$
defined by the relation $t_{\bi{Q}}^{\phd}f(\bi{k})=f(\bi{k+Q})$.
Relying on the commensurability property of the wave-vector,
$\bi{k}+2\bi{Q}=\bi{k}$, and the concomitant relation
$t_{\bi{Q}}^{\phd}\Delta(k,k+\bi{Q})=
\Delta(k+{\bi{Q}},k+2{\bi{Q}})=\Delta(k+\bi{Q},k)$, we can express
${\cal D}$ in the following manner

\bea{\cal D}(k)=\left(\begin{array}{cc}0&\Delta(k,k+\bi{Q})\\
t_{\bi{Q}}^{\phd}\Delta(k,k+{\bi{Q}})&0\end{array}\right)\,.\eea

\noi The latter enunciates that the form of ${\cal D}$ will be
determined by the behaviour of $\Delta$ under the action of
$t_{\bi{Q}}^{\phd}$. Decomposing $\Delta$ into periodic
$\Delta^+(k)$ and antiperiodic $\Delta^-(k)$ parts we obtain the
following concrete expression for ${\cal D}$ \bea {\cal
D}(k)=\Delta^{+}(k)\tau_{1}^{\phd}+\Delta^{-}(k)\rmi\tau_{2}^{\phd}\,.\eea

\noi The fact that ${\cal D}$ is a hermitian operator also implies
that \bea\Delta^+(k)&=&+[\Delta^+(k)]^{*}\Rightarrow \Delta^+(k)\quad{\cal R}eal\,,\\
\Delta^-(k)&=&-[\Delta^-(k)]^{*}\Rightarrow \Delta^-(k)\quad{\cal
I}maginary\,.\eea

\par Based on the preceding results we conclude that a
$\rm{d_{{xy}}}^{\phd}$ density wave is real whilst a $\rm{d}_{{\rm
x^2-y^2}}$ density wave is imaginary. As a result a $\rm{d}_{{\rm
xy}}$ component violates parity $(k_{{\rm x}}^{\phd},k_{{\rm
y}}^{\phd})\rightarrow(k_{{\rm x}}^{\phd},-k_{{\rm y}}^{\phd})$
while a $\rm{d}_{{\rm x^2-y^2}}$ component violates time reversal
because it is imaginary. This implies that the density wave is
chiral as it possesses angular momentum in $\bi{k}-$space
perpendicular to the ${\rm x-y}$ plane. This angular momentum acts
as an intrinsic magnetic field already present in the ground state
of the chiral density wave affecting dramatically its
electromagnetic properties. A crucial consequence of this
intrinsic magnetic field is that it permits the appearance of a
Hall voltage with the sole application of an electric field. This
phenomenon is named Spontaneous Quantum Hall Effect. Contrary to
the usual Quantum Hall Effect, here there is no need for an
external magnetic field for the phenomenon to occur (spontaneous
character). This necessity is fully settled by the intrinsic
angular momentum induced by the ${\cal P}-{\cal T}$ violation.


Our next and final step is to derive the Chern-Simons terms
governing the long wavelength response of the chiral density wave
leading to the Spontaneous Quantum Hall Effect. According to the
previous symmetry considerations, the chiral d-wave density wave
is described by an order parameter of the form
$\Delta(\bi{k})=\eta\Delta_0^{\phd}\sin k_{\rm{x}}\sin
k_{\rm{y}}+\rmi\Delta_{0}^{\phd}(\cos k_{\rm{x}}-\cos
k_{\rm{y}})$. $\Delta_0^{\phd}$ is the modulus of the imaginary
part while $\eta$ defines the relative magnitude of the two
components and its sign determines the direction of the chirality.
To obtain an effective action for the gauge field, we take into
account its fluctuations through its coupling with the electrons.
The corresponding action is

\bea\tilde{S}&=&\tilde{S}_{\rm{gs}}-\frac{\rmi}{2}\Tr\ln\left\{I-\hat{{\cal
G}}\hat{{\Gamma}}_{\mu}^{\phd}\hat{A}^{\mu}\right\}\no\\&=&
\tilde{S}_{\rm{gs}}^{\phd}+\frac{\rmi}{2}\Tr\sum_{n=1}^{\infty}\frac{1}{n}\left\{\hat{{\cal
G}}\hat{{\Gamma}}_{\mu}^{\phd}\hat{A}^{\mu}\right\}^n\,.\eea

\noi By expanding the logarithm, we managed to work out a
perturbation series for the electromagnetic part of the action. In
the lowest order approximation we have \bea
S_{\rm{em}}^{\phd}=\frac{1}{2}\int_{q}\ph
A^{\mu}(-q)\Pi_{\mu\nu}^{\phd}(q)A^{\nu}(q)\,,\eea

\noi where we have introduced the polarization tensor
$\Pi_{\mu\nu}$ defined by the equation

\bea\fl\Pi_{\mu\nu}^{\phd}(q)&=&\frac{\rmi}{2}\int_{\omega}\sum_{\bi{k}}\ph
Tr\left[{\cal G}(k)\Gamma_{\mu}^{(p)}(k){\cal
G}(k+q)\Gamma_{\nu}^{(p)}(k+q)\right]\no\\&-&\frac{\ph
e^2}{m}\rho_{\rm{e}}^{\phd}\delta_{i,j}^{\phd}\,,\eea

\noi where $\rho_{\rm{e}}^{\phd}$ is the electron density in two
dimensions, without taking into account the spin degrees of
freedom. The long wavelength treatment of the tensor yields
\bea \Pi_{00}^{\phd}(q)&=&{\cal O}(q^2)\,,\\
     \Pi_{0i}^{\phd}(q)&=&+\rmi\sigma_{\rm{xy}}^{\phd}\varepsilon_{0ij}^{\phantom{\dag}}q^j\,,\label{eq:P}\\
     \Pi_{i0}^{\phd}(q)&=&-\rmi\sigma_{\rm{xy}}^{\phd}\varepsilon_{0ij}^{\phantom{\dag}}q^j\,,\\
     \Pi_{ij}^{\phd}(q)&=&+\rmi\sigma_{\rm{xy}}^{\phd}\varepsilon_{i0j}^{\phantom{\dag}}q^0/2\,,\eea

\noi where we have introduced the Hall conductance $\sigma_{{\rm
xy}}^{\phd}$ and the totally antisymmetric symbol
$\varepsilon_{0ij}^{\phantom{\dag}}$ . We observe that
$\Pi_{00}^{\phd}(0)=\Pi_{ij}^{\phd}(0)=0$ in accordance with gauge
invariance. The terms
$\Pi_{0i}^{\phd},\Pi_{i0}^{\phd},\Pi_{ij}^{\phd}$ provide the
Chern-Simons action that we have already mentioned. In real space,
this action has the form \bea S_{CS}^{\phd}=\frac{\sigma_{{\rm
xy}}^{\phd}}{4}\int_x\ph
\varepsilon_{\mu\nu\lambda}^{\phantom{\dag}}A^{\mu}F^{\nu\lambda}\,,\eea

\noi where $\varepsilon_{\mu\nu\lambda}$ is the totally
antisymmetric symbol and
$F_{\mu\nu}=\partial_{\mu}A_{\nu}-\partial_{\nu}A_{\mu}$. This
action is responsible for mixing the magnetic and electric fields
leading to a Hall Effect. In fact this action appears in the usual
Quantum Hall Effect where an external electric and an external
magnetic field are present. The quantum character of the
phenomenon relies on the topological structure of the Hamiltonian.
\par In the case we are studying, the ${\cal P-T}$ violation induced
by the chiral d-density wave is responsible for the generation of
the Chern-Simons action. This implies that in our case the system
exhibits a (reciprocal) Hall response with the sole application of
an external (magnetic) electric field, justifying in this manner
its spontaneous character. Consequently, a chiral d-density wave
exhibits the Spontaneous Quantum Hall Effect described by the
following equation \bea j_{i}^{\phd}=\sigma_{{\rm
xy}}^{\phd}\varepsilon_{0ij}^{\phd}E_j^{\phd}\,\eea

\noi where $E^{j}$ is the $j-th$ component of the electric field.
As far as the quantum character of the phenomenon is concerned,
this relies once again on the topological aspects of the ground
state. We compute the Hall conductance using eq.~(\ref{eq:P})
obtaining
\bea\sigma_{{\rm{xy}}}^{\phd}&=&\frac{\rmi}{2!}\varepsilon_{0ji}\frac{\partial\Pi_{0i}^{\phd}}{\partial
q^j}=\frac{e^2}{16\pi^2}\int{\rm dk}_{{\rm x}}{\rm dk}_{\rm
y}\ph\hat{\bi{g}}\cdot\left(\frac{\partial\hat{\bi{g}}}{\partial
k_{\rm{x}}}\times\frac{\partial\hat{\bi{g}}}{\partial
k_{\rm{y}}}\right)\no\\&=&\frac{e^2}{4\pi}\widehat{N}\,,\eea

\noi where we have set
$\hat{\bi{g}}={\bi{g}}/{\mid\bi{g}\mid}={\bi{g}}/{E}$ and
introduced the topological invariant
\bea\widehat{N}=\frac{1}{4\pi}\int {\rm d^2k}\phd
\hat{\bi{g}}\cdot\left(\frac{\partial\hat{\bi{g}}}{\partial
k_{\rm{x}}}\times\frac{\partial\hat{\bi{g}}}{\partial
k_{\rm{y}}}\right)\,,\eea

\noi which is an integer that performs a mapping of a sphere to
another sphere. We have to explain what are the spheres in our
case. Eq.~(\ref{eq:s}), points out that the behaviour of the
system is determined by the orientation of the $\bi{g}$ vector in
isospin space. The finiteness of energy imposes the vanishing of
the order parameter at the boundary. This means that at the
boundary, $\bi{g}$ has a fixed orientation along the $3^{rd}$
isospin space axis. The fixed orientation makes the boundary
points equivalent. In this manner, we can replace the whole
boundary by one point, making our two dimensional space a sphere.
The other sphere is the order parameter's configuration space.
These two spheres are mapped to each other via $\widehat{N}$. In
our case $\widehat{N}=2$ because the order parameter components
are eigenfunctions of the angular momentum in momentum space with
$l=2$. The Hall conductance is analogous to $\widehat{N}$. So we
conclude that $\sigma_{{\rm xy}}^{\phd}$ is a topological
invariant and is equal to (per one spin component)
\bea\sigma_{{\rm xy}}^{\phd}=\frac{e^2}{2\pi}\eea

\noi We have to remark that in order to obtain the value of the
bulk Hall conductance we should multiply the latter result with
the thickness of the material along the z-axis.


In conclusion, we studied the unusual electromagnetic behaviour of
a chiral $\rm{d_{xy}+id_{x^2-y^2}}$ density wave. The ${\cal P-T}$
violation by this ground state is responsible for the generation
of an intrinsic angular momentum that plays the role of an
intrinsic ``magnetic'' field. As a consequence, the application of
an electric field suffices to obtain a Hall response, leading this
way to a Spontaneous Hall Effect. The Hall conductance is
quantized as a topological invariant reflecting the presence of
the internal angular momentum in the ground state. We argue that
the SQHE state is observable and may constitute the fingerprint of
a chiral d-density wave state. Since a $\rm{d_{x^2-y^2}}$ state
has been so often proposed for so many different materials, close
to half filling is quite natural to expect the emergence of such
chiral states. Finally, we have to remark that similar results are
expected in the case of a $\rm{d_{xy}+id_{x^2-y^2}}$ spin triplet
density wave state in the case in which the two components share
the same spin polarization.

\ack

The authors are grateful to Professor P. B. Littlewood for
valuable advice and suggestions. Moreover, G.V. acknowledges
motivating discussions with J. Goryo on the Spontaneous Quantum
Hall Effect during a stay at Dresden supported by P. Thalmeier and
the MPI CPfS. This work has been supported by the EU STRP grant
NMP4-CT-2005-517039. P.K. also acknowledges financial support by
the Greek Scholarships State Foundation.

\end{document}